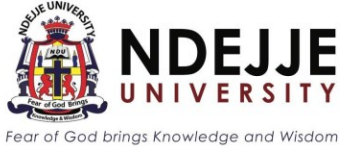

# New Curriculum, New Chance – Retrieval Augmented Generation for Lesson Planning in Ugandan Secondary Schools.
Prototype Quality Evaluation.

**Working Paper**

Simon Kloker
Ndejje University, Coworkers
skloker@ndejjeuniversity.ac.ug

Herbertson Bukoli
Ndejje University
hbukoli@ndejjeuniversity.ac.ug

Twaha Kateete
Ndejje University
tkateete@ndejjeuniversity.ac.ug

## ABSTRACT

**Introduction:** Poor educational quality in Secondary Schools is still regarded as one of the major struggles in 21st century Uganda – especially in rural areas. Research identifies several problems, including low quality or absent teacher lesson planning. As the government pushes towards the implementation of a new curriculum, exiting lesson plans become obsolete and the problem is worsened. Using a Retrieval Augmented Generation approach, we developed a prototype that generates customized lesson plans based on the government-accredited textbooks. This helps teachers create lesson plans more efficiently and with better quality, ensuring they are fully aligned the new curriculum and the competence-based learning approach.

**Methods:** The prototype was created using Cohere LLM and Sentence Embeddings, and LangChain Framework – and thereafter made available on a public website. Vector stores were trained for three new curriculum textbooks (ICT, Mathematics, History), all at Secondary 1 Level. Twenty-four lessons plans were generated following a pseudo-random generation protocol, based on the suggested periods in the textbooks. The lesson plans were analyzed regarding their technical quality by three independent raters following the Lesson Plan Analysis Protocol (LPAP) by Ndihokubwayo et al. (2022) that is specifically designed for East Africa and competence-based curriculums.

**Results:** Evaluation of 24 lesson plans using the LPAP resulted in an average quality of between 75 and 80%, corresponding to "very good lesson plan". None of the lesson plans scored below 65%, although one lesson plan could be argued to have been missing the topic. In conclusion, the quality of the generated lesson plans is at least comparable, if not better, than those created by humans, as demonstrated in a study in Rwanda, whereby no lesson plan even reached the benchmark of 50%.

**INTRODUCTION**

Poor educational quality in Ugandan secondary schools, particularly in rural areas, remains a major challenge in 21st century (Kakuba et al., 2021; Lucas et al., 2014; Malunda et al., 2016; Naamara et al., 2017). Research identifies low-quality or absent teacher lesson planning as prominent cause, among many more (Businkye & Najjuma, 2015; Guloba et al., 2010; Malunda et al., 2016; Mpaata & Mpaata, 2019; Okiror et al., 2017). Lack of materials, lack of time, or lack of training are mentioned as reasons therefore. However, the evidence that existent and good lesson plans increase quality of education significantly is strong (Bagaya et al., 2020; Omara et al., 2021; O'Sullivan, 2006). While school inspections haven't shown significant positive effects (Bagaya et al., 2020) school-based interventions promoting creativity and innovativeness, as well as providing materials have been successful (Omara et al., 2021).

The recent introduction of a new curriculum aiming to better align with global standards and address diverse student needs signifies the government's commitment to improving education. This reform is crucial for teachers, students, and the government as it seeks to improve the overall quality and relevance of education (Kidega et al., 2024). Personalized and well-structured lessons can enhance student engagement and learning outcomes (Ashcraft, 2024). However, the implementation of the new curriculum poses significant challenges, particularly in the area of lesson planning (Kidega et al., 2024). Teachers often struggle with limited access to teaching materials, time constraints, and the need for personalized instructional strategies, which can hinder their ability to deliver effective lessons (Iqbal et al., 2021; Kidega et al., 2024). The problems increase, as the implementation of this new curriculum makes exiting lesson plans obsolete. Without adequate support and resources, teachers may find it difficult to fully realize the curriculum's potential, thereby limiting its impact on student learning (Kidega et al., 2024).

Large Language Models (LLMs) have emerged as a potential solution to address lesson plan creation challenges. Generative Artificial Intelligence (GenAI) as another often-used term holds the potential to transform lesson planning by automating and enhancing the process through analyzing curricula and student data to generate (and refine) dynamic, customized lesson plans (Baytak, 2024; Karpouzis et al., 2024; Kehoe, 2023; Lodge et al., 2023; Sakamoto et al., 2024). However, Karpouzis et al. (2024) argues that this technological progress comes with challenges. One major concern is the potential overdependence on technology for lesson planning. There's a risk that educators may rely too heavily on AI tools, which could undermine their role in the creative and professional aspects of lesson planning. Also, education accuracy and to generic outcomes are mentioned. Other authors mention unpredictability of results, as well and inconsistency with educational approaches or contextual factors (such as class sizes) as a problem (Baytak, 2024; Sakamoto et al., 2024). Educators must retain a significant role to ensure AI enhances, not replaces, their expertise, context awareness and creativity.

Retrieval Augmented Generation (RAG) is an advanced approach that combines retrieval-based and generation-based models to produce high-quality, contextually relevant content. By leveraging RAG, context and tailored content can be injected in the generation process. This information is retrieved form government-approved textbooks and supports the LLM to stick to contextualized, pedagogically sound frameworks, thereby enhancing their instructional capabilities and efficiency. The approach comes with advantages both for teachers and the government. Teachers can generate new lesson plans more effectively, while the new curriculum with its new topics and competence-based teaching approaches is promoted. The innovativeness and creativity of the curriculum developers, text-book authors and the LLM are all considered and potentially help teachers to select the most fitting approach. For students, the implementation of RAG in lesson planning translates to improved learning experiences.

Therefore, we propose the use of RAG in lesson planning. This project focuses on developing a prototype software tool that generates lesson plans based on government-approved textbooks and adheres to the suggested layout. Our research question is: "Does the quality of lesson plans generated by RAG using government-approved textbooks meet necessary standards for implementation in Ugandan classrooms?"

## RELATED LITERATURE

### Uganda's Educational Quality and the Role of Lesson Planning

Studies reveal consistently low academic performance among primary and secondary students in developing countries like Uganda, as evidenced by teacher assessments (Abenawe, 2022; Kjær & Muwanga, 2019). The literature identifies several potential causes, including limited teacher engagement, absenteeism, unfavorable student-teacher and classroom-pupil ratios, and a lack of teaching materials.

The Ugandan Government is actively addressing these issues through various interventions. These initiatives include establishing teacher management information systems, standardizing teacher training programs, and continuous professional development by integrating ICT into pedagogical skills training across teacher training institutions (Komakech & Osuu, 2014). Additionally, the government introduced a new competency-based curriculum for secondary schools in 2008, replacing the knowledge-based approach and revising content areas (Olema et al., 2021). Implementation began in 2022 for lower-level secondary schools, with new textbooks included.

However, as in other developing countries, efficient curriculum implementation remains a challenge in Uganda. Teachers often prioritize exam preparation over curriculum content (Barasa et al., 2023; Olema et al., 2021). A significant concern regarding the new curriculum is the increased workload for teachers, which may not consider their professional, financial, and personal circumstances (Barasa et al., 2023). Part of this increased workload is the revision of existing lesson plans or the creation of new ones.

Uganda National Examinations Board released a report in 2014, showing evidence that most teachers lack the necessary skills for effective lesson planning and teaching. This finding aligns with research from Uganda and East Africa more broadly (Apolot et al., 2018; Komakech & Osuu, 2014; Malunda et al., 2016; Ndihokubwayo et al., 2020). In many Ugandan secondary schools, teachers rely on rudimentary schemes of work and outdated notes, with minimal or poorly designed lesson plans, ultimately hindering the long-term quality of education (Bagaya et al., 2020).

### Lesson Planning in Uganda

A lesson plan serves as a blueprint for effective instruction, guiding teachers in their classroom practice. It's a preparatory activity undertaken before implementation in a specific classroom setting (Raval, 2013; Savage, 2015). Key elements include learning objectives, learning outcomes, teaching activities, resources, materials, differentiation strategies, and assessment strategies (Savage, 2015). Lesson plans embody a teacher's decisions for effective preparation (Taylan, 2018) and it is a professional responsibility or teachers to prepare them for every lesson they teach (Omara et al., 2021).

Habyarimana et al. (2017) investigated teachers' self-beliefs on their effort and ability in Ugandan Public-Private Partnership schools. Their findings revealed an average daily teaching time of only 2 hours and 20 minutes. Notably, only 89% of the teachers claimed to have prepared schemes of work and only 67% claimed to have prepared lesson plans for a week. his suggests a potential preference for schemes of work over lesson plans among secondary school teachers, as further supported by Bagaya et al. (2020). Time constraints appear to be a major factor in the lack of robust lesson plans. Roberts (2015) highlights a trade-off between class attendance and lesson planning in a Tanzanian context, suggesting that the time required for lesson planning contributes to teacher absenteeism.

Teacher capability in crafting strong lesson plans is another concern. In-service and preparatory training shows mixed effects, with a potential positive impact if well-designed. While Komakech & Osuu (2014) report negligible differences in lesson planning between trained and untrained Ugandan teachers, Omara et al. (2021) found that professionally supervised student teaching practice enhances lesson preparation, planning, presentation, and delivery. Same is reported by Giacomazzi et al. (2023), that developed a model and training for critical thinking for teachers in Uganda. Malunda et al. (2016) and Saoke et al. (2022) further advocate for increased in-service training to improve lesson planning in Uganda and Kenya, respectively. Government supervision and headteacher monitoring also demonstrate positive impacts on lesson planning (Malunda et al., 2016; Mugabe & Ogina, 2021).

Beyond time constraints and capability, some sources report a lack of awareness regarding the significance of lesson plans for quality education. Many teachers prepare lesson plans just for purposes of evaluation by the school administration (Sawyer & Myers, 2018; Theoharis & Causton-Theoharis, 2011) neglecting the potential for well-prepared lesson plans to serve as substitutes in their absence (Jacob et al, 2008).

**Lesson Plan Generator**

Early research focused on supporting lesson creation, not generation. Strickroth (2019) reviewed existing lesson planning systems and compared eleven of them, finding none to meet all predefined requirements by his study. Consequently, they developed their own tool, PLATON. While these tools vary in support depth and guidance style, they primarily provide teachers with forms that still require manual completion.

Their benefits lie in prompting reflection for inexperienced teachers and offering best-practice guidance (Strickroth, 2019). However, they do not provide actual content (and therefore the quality of content of the lesson plans in their study have not been evaluated).

More recent work explores and evaluates the (iterative) generation of complete lesson plans, including content, that teachers can then adapt. Baytak (2024) compared Google Gemini and OpenAI's ChatGPT for lesson plan creation. The study found these large language models (LLMs) generate lesson plans resembling human-made ones, but with limitations: limited technology integration, high dependence on users' ability to make up good prompts, and limited understanding of the context. These latter issues were also identified by Sakamoto et al. (2024) in both Japanese and Canadian contexts: Creating good lesson plans using LLMs requires multiple prompts, analysis, and iterations to align with curriculum and also context. Karpouzis et al. (2024) compared Google Bard, ChatGPT, and Llama, reaching similar conclusions. Also, they mention that the generated lesson plans may lack accuracy in the specific educational settings, meaning not sticking to the relevant teaching approaches or not considering the context sufficiently. However, van den Berg & du Plessis (2023) emphasize the potential of LLMs to enhance teacher creativity, critical thinking, and openness during lesson planning.

**Methodology**

The overall research project follows a two-cycle Design Science Research approach (Vaishanavi & Kuechler, 2004), entering on the NDU Lesson Plan Generator (NLPG) artifact. The current describes the first design cycle, which focuses on the technical evaluation of the generated lesson plans.

The technical core of the NLPG follows standard approaches from Retrieval Augmented Generation. Information for given topics is retrieved from government accredited textbooks and provided to a Large Language Model for further processing to the desired output. The textbooks have been provided as .pdf files and pre-processed to vector stores using Cohere Text Embeddings[1]. The results were exported to the server of the retrieval system (web application). The web application itself retrieves the relevant vector store for the selected subject and uses it within a LangChain[2] retriever chain, combined with the Cohere LLM. The LLM is instructed to only use the information provided from the retriever chain to base its lesson plans on, thereby ensuring the lesson plan adheres to the new curriculum and government-approved information. The LLM is also instructed to output the result in the lesson plan format suggested by the government and thereby, among others, stick to the competence-based teaching approach.

---

[1] https://cohere.com/embeddings

[2] https://python.langchain.com/v0.1/docs/modules/data_connection/retrievers/

The prototype was trained on three textbooks, all secondary one level for the subjects History (& Political Education), Mathematics, and ICT. For History and Mathematics, we used the student's version of the textbooks, while for ICT we used the teacher's version of the textbooks.

The web application itself is hosted on a python web server and employs the Flask framework. The interface is fully created according to principles of responsive design and communicates asynchronously to the server for improved user experience. It was also designed to minimize data consumption by avoiding large image files or elaborate animations, considering that the proposed user base consists of teachers in rural areas as well. Figure 1 shows the lean interface of the resulting web application.

To evaluate the quality of the lesson plans we followed the approach of Ndihokubwayo et al. (2022). We slightly adapted their published Lesson Plan Analysis Protocol (LPAP) by matching wording of Rwandan curriculum to Ugandan curriculum and suggested lesson plan format (e.g. "Title" → "Topic", "Key Unit Competence" → "Learning Objective", or "Conclusion" → "Wrap-up and Assessment") and additionally dropped five of the 27 six items as they are either not applicable to the Ugandan lesson plan format (e.g. the Ugandan lesson plan format does not specifically outline instructions for special education needs or instructional objectives) or to our use case (e.g. "Lesson Evaluation", as our lesson plans have not been taught). The scoring table has been adapted accordingly to 44 instead of 54 maximum points.

We generated 24 lesson plans, eight for each subject. We selected the topics in a quasi-random manner by picking every second topic in table of contents of the textbooks. If a topic was obviously too broad, we selected the first subtopic. For consistency, we constantly chose a class size of "> 60" and for number of periods always "1". If the generated lesson plan exhibited incorrect HTML formatting (not content-related issues), we repeated the generation.

Subsequently, three independent raters evaluated the 24 lesson plans according to a pre-defined protocol. We ensured the raters had a common understanding of the necessary terminology. All raters possessed multiple years of teaching experience and were proficient in English. In some instances, the NLPG generated multiple periods, although we instructed it to generate only one. In these cases, we solely included the first period in the evaluation and disregarded any subsequent periods.

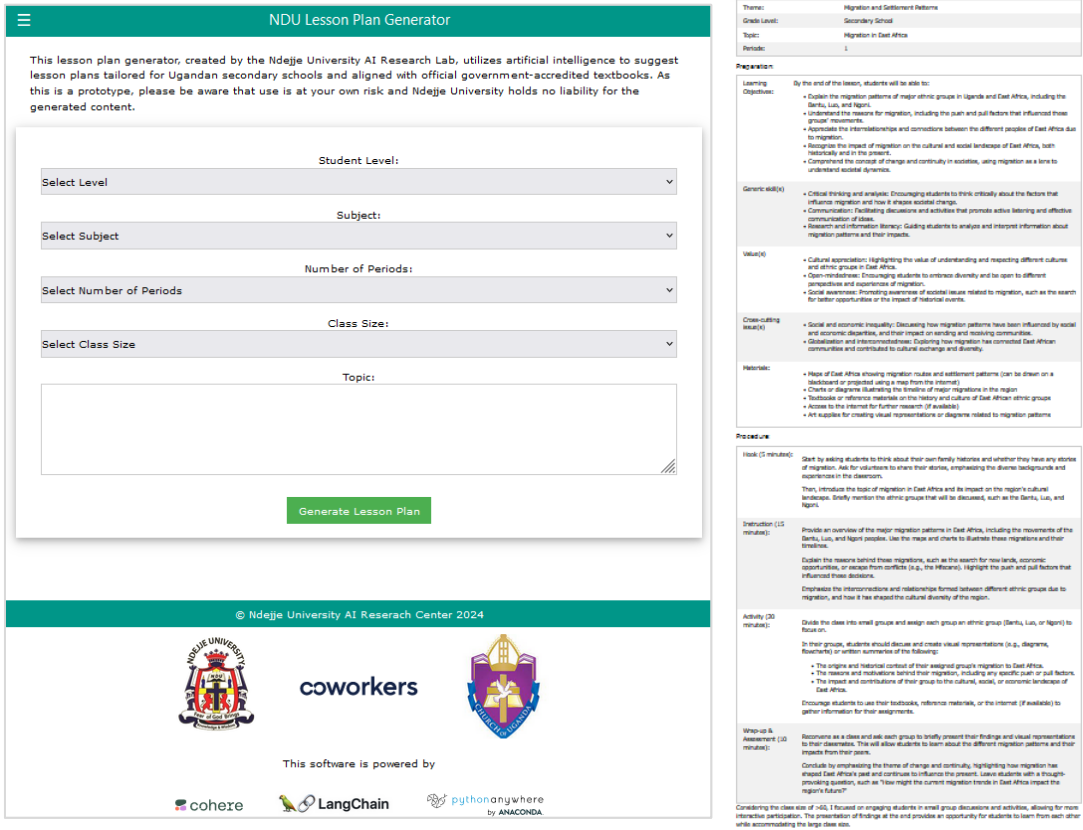

**Figure 1** Left: Interface of the NDU Lesson Plan Generator. A lean form contains four dropdowns to select student level, subject, number of periods and class size. A free-text field allows to specify the topic of the lesson plan. Right: A resulting lesson plan in the Ugandan lesson plan format divided in three sections: General Information, Preparation, and Procedure.

We employed similar rating quality measures as Ndihokubwayo et al. (2022) and averaged the three independent ratings on a value basis for scoring. The full dataset and the 24 generated lesson plans can be downloaded from the online appendix.

**RESULTS**

Figure 2 depicts the inter-rater agreement using metrics analogous to Ndihokubwayo et al. (2022), including percent agreement, Spearman's Correlation, and Kappa Coefficient. Our scores are generally comparable to those reported in Ndihokubwayo et al. (2022). However, we observe a slightly higher number of outliers, particularly towards the lower end of the agreement scale. This discrepancy is likely attributable to a single lesson plan (*History07.pdf*) that demonstrably did not address the intended topic, leading to greater room for rater interpretation.

For lesson plan *History07.pdf*, the assigned topic was "Kenya". However, the corresponding textbook section on this topic comprised less than a single full page. Consequently, the generated lesson plan addressed "Mukasa" instead of Kenya. Although the lesson plan showed good teaching approaches, the topic was still wrong. This is reflected in the overall scores presented in Figure 3. The figure also illustrates that the average scores for lesson plans in Mathematics and ICT are nearly identical, while History exhibited a higher score. If lesson plan *History07.pdf* is excluded due to the topic discrepancy, the difference even becomes significant at a 5% level (Wilcoxon signed-rank test).

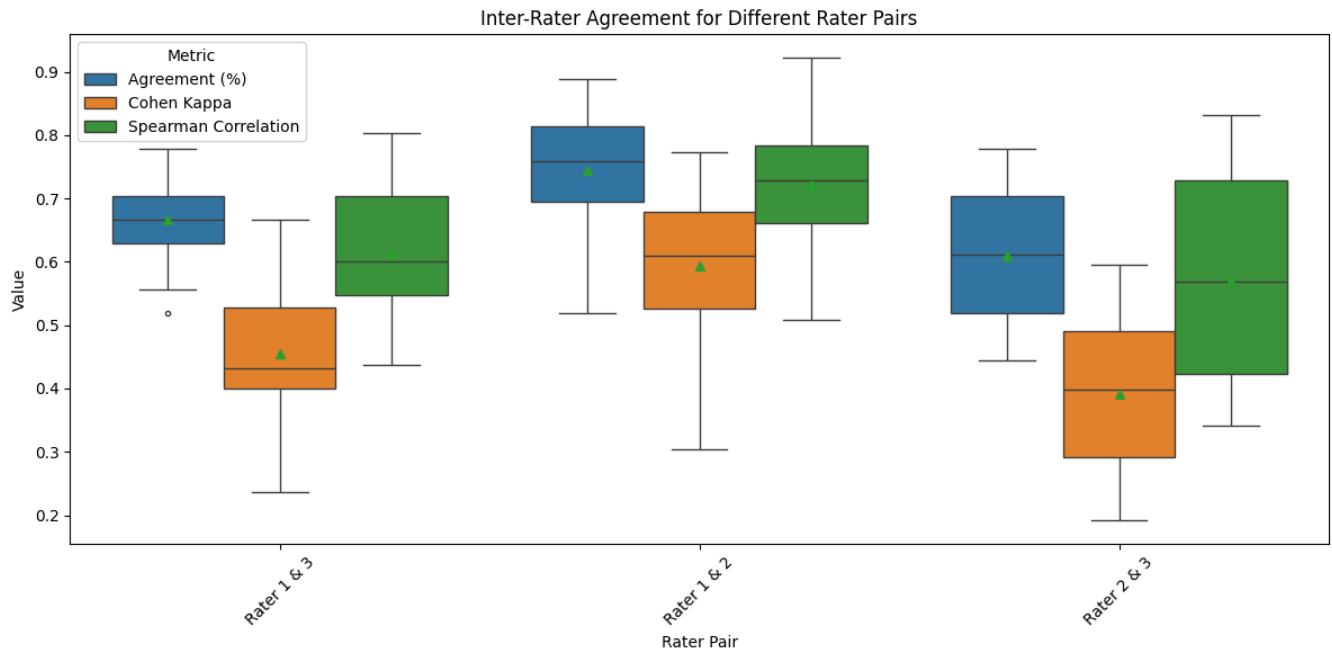

**Figure 2 Inter-rater agreement in box-plots.**

More importantly, our lesson plans consistently outperform the human-created lesson plans reported by Ndihokubwayo et al. (2022). None of the lesson plans in Ndihokubwayo et al. (2022) achieved a score exceeding 50%. This is, to a certain extent, unsurprising. The NLPG is inherently advantaged in some aspects. Since it adheres to the Ugandan lesson plan format, it cannot omit mandatory sections in the final lesson plan. Items within the LPAP that assess the mere presence of these sections will invariably receive full credit. However, even when we restrict our evaluation to elements subject to qualitative variation, the average score based solely on these columns remains at 75%. According to Ndihokubwayo et al. (2022) this would still qualify as a "good lesson plan". Scores exceeding 80% are considered "very good" or "excellent" (above 90%). Therefore, we can conclude that the NLPG generates lesson plans that fulfill the necessary (technical) quality criteria to be implemented directly in classrooms (without requiring human adaptation, although this is still recommended) and demonstrably surpass the quality of human-created lesson plans referenced in other LPAP-based studies.

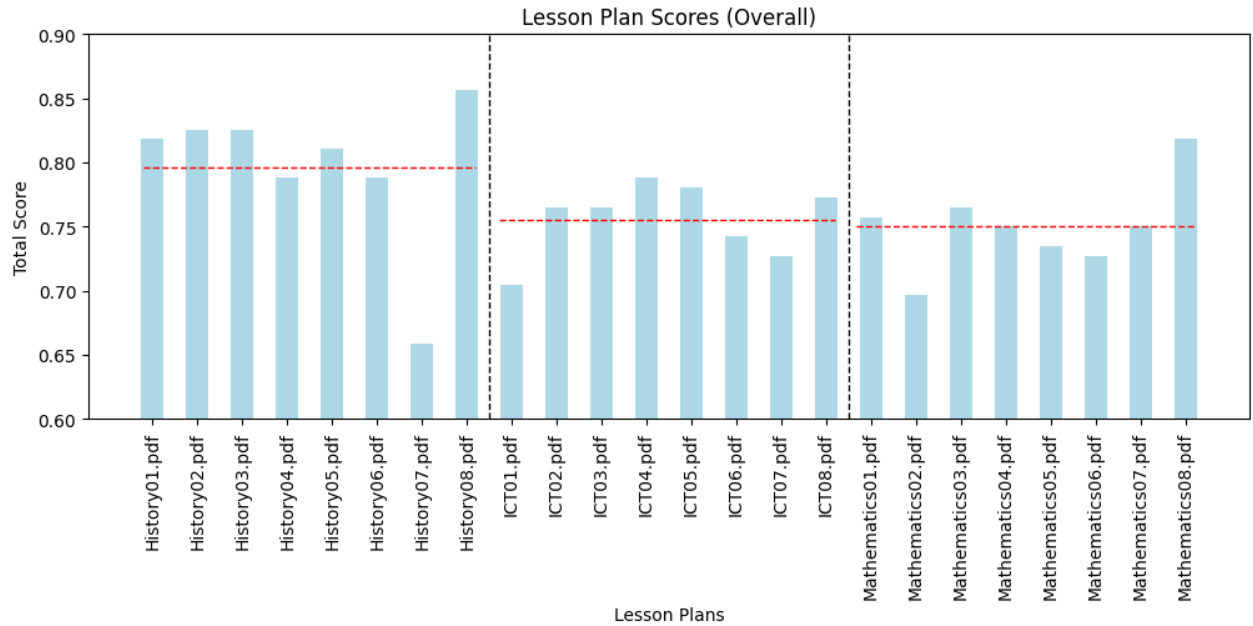

Figure 3 Overall Scores of the lesson plans according to the LPAP.

## DISCUSSION

Previous works on LLMs for lesson planning reported lesson plan quality at least competitive to human created lesson plans (Karpouzis et al., 2024; Sakamoto et al., 2024). However, these studies have been conducted in northern hemisphere. Our results, in the context of East African developing countries (i.e. Ndihokubwayo et al., 2020, 2022), even exceed the quality of human made lesson plans. There are likely two reasons for this. First the human made lesson plans in East African context might be of lower quality than those in high income countries due to worse education or working conditions. Second, and more importantly, our approach avoids producing "generic" content. It bases the results on contextualized Ugandan textbooks, already considering relevant content and teaching approaches.

There have been several other ways to provide teachers with such contextualized lesson plans in the past, such as those delivered with teacher's textbooks or provided by the government or publishers. However, it's important to note that these pre-made plans lack both the randomness and context-specificity of our solution. The NLPG generates a different lesson plan each time, even with no parameter changes. This fosters teachers' critical thinking and exposure to a variety of methods and approaches. Secondly, only the teacher has complete knowledge of the exact class and context. A future version of the NLPG should allow teachers to set even more parameters (beyond class size) to tailor the resulting lesson plan further.

A crucial current limitation is the system's hallucination problem, where the LLM invents facts when lacking sufficient knowledge. This often coincides with the "confidence conundrum": the LLM presents information confidently regardless of its actual certainty. Two chapters in Searson et al. (2024) discuss LLM hallucination as well in the context of LLM in education, but none of the sources focusing on LLM discuss this issue yet in the context of lesson planning. Lesson plan *History07.pdf* is an example where our model fabricated facts, as it did not have enough information about "Kenya" (the requested topic) in the underlying textbook (not even one full page). Knowledgeable and diligent teachers will quickly identify the suspicious and faulty lesson plan. However, busy or unprepared teachers might risk teaching fabricated information. Equally concerning, this behavior undermines teachers' acceptance of and trust in the tool and its perceived usefulness. Possible countermeasures that should be considered in future versions are displaying information about the LLMs confidence (e.g., number of retrieved pages in the textbooks) or crosschecking topic and/or result against plausibility with the retrieved information.

Several other limitations require discussion: (i) The study only covered three subjects for now. We chose them for content type variety, but it remains a limited sample. (ii) We only evaluated technical quality. The content quality itself was not rated by teachers with everyday classroom experience, but by university lecturers. (iii) The used textbooks were prototypes from the National Council for Higher Education (NCHE), not publisher versions. We don't expect significant differences, but wanted to acknowledge this. Further issues, that have not yet been discussed are ethical considerations and social implication, such as if the introduction of GenAI in the teacher-student interaction undermines the trust-relationship (i.e. Kloker et al., 2024).

**CONCLUSION**

This research investigated the potential of RAG for lesson plan creation in the context of Ugandan secondary schools. We developed a prototype, that utilizes an LLM to generate lesson plans based on government-approved textbooks aligned with the new curriculum. The quality of the generated lesson plans was evaluated using the LPAP by Ndihokubwayo et al. (2022), adapted for the Ugandan context.

The results are promising. The NLPG-generated lesson plans achieved an average score of 75-80%, exceeding the quality of human-created lesson plans reported in similar studies (none exceeding 50%). The inter-rater agreement scores were comparable to those reported by Ndihokubwayo et al. (2022), indicating reliable evaluation. This suggests that the NLPG can generate lesson plans that meet the necessary standards for implementation in Ugandan classrooms and technical quality is assured.

However, limitations exist. The system is right now susceptible to LLM hallucination, where it invents facts in cases there is insufficient knowledge in the textbooks. Lesson plan "History07.pdf" exemplifies this, where the model lacked enough information about "Kenya" and fabricated content. Additionally, the study only covered three subjects, evaluated technical quality, and used prototype textbooks.

Future research should address these limitations. As the current prototype evaluation is part of a multi cycle design science project, it contributes valuable information to further develop the artifact into a useful and novel contribution both in academia and secondary schools. Expanding the subject range, incorporating teacher evaluations of content quality, and utilizing official textbooks are crucial next steps. The evaluation scheme from Strickroth (2019) might be a promising tool to add these insights. Furthermore, exploring methods to mitigate LLM hallucination, such as confidence scores or plausibility checks, is essential. Interviews with teachers and student teachers should also increase insight in how the tool might be used exactly and what are necessary features to be accepted and used.

Other next steps, not part of the current project, would be e.g., to incorporate recent findings in how to make lesson plans more effectively in general. König et al. (2022), i.e., demonstrates that ICT integration in teachers' lesson plans result in better learning outcomes.

Overall, this research demonstrates the potential of RAG for generating high-quality lesson plans in Ugandan secondary schools. By addressing the limitations and continuing development, the NLPG can be a valuable tool to support teachers, improve lesson planning quality, and ultimately enhance student learning outcomes. Additionally, it can contribute in the government's efforts to deploy the new curriculum faster and more efficiently as well as it is interesting to publishers that want to increase their textbooks values by accompanying it by tailored lesson plan creation services.

**APPENDIX**

Online Appendix can be accessed here: https://drive.google.com/drive/folders/1DPK-PyDr-Mb1H9F_LjeP_ie59SIy5kiV?usp=sharing